\documentclass[a4paper,11pt]{article}
\usepackage{amsmath,amssymb,amsfonts,indentfirst,graphicx,color,anysize,hyperref}
\marginsize{2.5cm}{2.5cm}{2cm}{2.0cm}
\graphicspath{{./figs/}}
\usepackage{etoolbox}
\apptocmd{\thebibliography}{\raggedright}{}{}

\title{Femtoscopy with Lévy sources from SPS through RHIC to LHC}

\author{M. Csanád, D. Kincses\\
	ELTE Eötvös Loránd University, Institute of Physics, \\Pázmány Péter s. 1/a, Budapest, H-1117, Hungary
}

\begin{document}

\maketitle

%%%%%%%%%%%%%%%%%%%%%%%%%%%%%%%%%%%%%%%%%%
\begin{abstract}
Femtoscopy is a unique tool to investigate the space-time geometry of the matter created in ultra-relativistic collisions. If the probability density distribution of hadron emission is parametrized, then the dependence of its parameters on particle momentum, collision energy, and collision geometry can be given. In recent years, several measurements came to light that indicated the adequacy of assuming a Lévy-stable shape for the mentioned distribution. In parallel, several new phenomenological developments appeared, aiding the interpretation of the experimental results or providing tools for the measurements. In this paper, we discuss important aspects of femtoscopy with Lévy sources in light of some of these advances, including phenomenological and experimental ones.
\end{abstract}

%%%%%%%%%%%%%%%%%%%%%%%%%%%%%%%%%%%%%%%%%%
\section{Introduction}

In high-energy physics, femtoscopy~\cite{Lednicky:2001qv,Lisa:2005dd} is the most important way to gain information about the space-time geometry of the particle-emitting source (the probability density of particle creation in an infinitesimal interval in phase-space). Historically, the phenomenon on which this technique is based was discovered by R. Hanbury Brown and R. Q. Twiss in astronomy in two-photon correlations~\cite{HanburyBrown:1952na}. Hence, measurements in this field are often labeled ``HBT-correlations''. Independently, G. Goldhaber and collaborators investigated proton--proton collisions and observed the correlations of identical pions~\cite{Goldhaber:1960sf}. Our current understanding suggests that both phenomena are based on the wave-like nature of light and pions, or alternatively (starting from the quantum nature of particle detection) on the Bose--Einstein symmetrization of pair wavefunctions. See, for example, the introduction of Refs.~\cite{Csanad:2018vgk,PHENIX:2017ino} for details.

Two-particle momentum correlations (or Bose--Einstein correlations) are defined as
\vspace{-6pt}
\begin{align}\label{e:c2def}
C_2(p_1,p_2)=\frac{N_{12}(p_1,p_2)}{N_1(p_1)N_2(p_2)},
\end{align}
where $p_1$ and $p_2$ are the momenta of the two particles, $N_{12}$ is the invariant pair-momentum distribution, while $N_1$ and $N_2$ are the invariant momentum distributions of the particles $1$ and $2$ (for identical particles, these distributions are of course also identical). In general, there are several reasons for correlated particle production (i.e., $C_2\neq 1$): for example, collective flow, jets, resonance decay, and conservation laws. In the case of high-multiplicity heavy-ion collisions, the main source of correlations is the HBT effect mentioned above, and they appear mostly at small relative momenta, where $p_1\approx p_2$, or if one introduces $K=(p_1+p_2)/2$ and \mbox{$q=p_1-p_2$}, the criterion $q\ll K$ may be given. It turns out that in the absence of final-state interactions and multiparticle-correlations, a simple relation can be established~\cite{Kurgyis:2020vbz,Nagy:2023zbg,Korodi:2022ohn} between the measured momentum correlation $C_2(q,K)$ and the spatial pair-distribution (or pair source) $D(r,K)$ (indicating the probability density of creating a pair in an infinitesimal interval around distance $r$ and mean momentum $K$), given as
\begin{align}\label{e:c2dr}
C_2(q,K)=1 + \int dr \exp (iqr) D(r,K) = 1 + \widetilde{D}(q,K),
\end{align}
where $\widetilde{D}$ denotes the Fourier-transform of the pair source $D(r,K)$ in its first variable (see a comprehensive overview of this result in Ref.~\cite{Nagy:2023zbg}). It is characteristic to sources in heavy-ion collisions that the dependence of $\widetilde{D}(q,K)$ and, hence, of $C_2(q,K)$ on $K$ is smoother than on $q$; this property is often called the ``smoothness approximation''~\cite{Lisa:2005dd}. One may furthermore note that several phenomena may decrease the strength of the correlation functions: decays of long-lived resonances~\cite{Csorgo:1994in}, coherence~\cite{Csorgo:1999sj}, and quasi-random electromagnetic fields~\cite{Csanad:2020qtu}. Taking these into account, the above equation is modified as
\vspace{-3pt}
\begin{align}\label{e:c2lambda}
C_2(q,K) = 1 + \lambda \cdot \widetilde{D}(q,K),
\end{align}
where $\lambda$ is the correlation strength parameter in the core--halo model~\cite{Csorgo:1994in} identified with the squared fraction of primordial pions amongst all produced pions.

Suppose one aims to characterize the geometry of the (pair) source. In that case, one often assumes a parametrization of it, in such a way that the source shape is a parametric distribution as a function of $r$, and its parameters depend on the pair momentum $K$. A usual assumption (see e.g.,~\cite{STAR:2020dav}) is that of a Gaussian source, in which case
\vspace{-6pt}
\begin{align}
D(r,K)\propto e^{-\frac{r^2}{R(K)^2}},
\end{align}
where $R(K)$ is the momentum-dependent width of the source (here, a factor of $2$ would appear in the denominator for the single-particle source, which then cancels from the pair source), which can also be called the HBT or femtoscopic size. This would lead to a correlation function of
\vspace{-6pt}
\begin{align}
C(q,K) = 1 + \lambda(K)\cdot e^{-q^2 R(K)^2}.
\end{align}

This HBT radius is of high relevance in terms of the physics of the quark--gluon plasma, as (especially if measured in multiple dimensions) it sheds light on transverse expansion or details of the phase diagram, as indicated, e.g., in Refs.~\cite{Lacey:2014wqa,STAR:2014shf,PHENIX:2014pnh,Lokos:2016fze,Cimerman:2017lmm}. We note here that, usually instead of pair momentum $K$, pair transverse mass $m_T=\sqrt{K_T^2 + m^2}$ is defined (in $c=1$ units), where $K_T$ is the transverse component of $K$, and $m$ is the mass of the particle \mbox{being investigated.}

Note finally that the above is true for non-interacting particles. However, even pions undergo electromagnetic final-state interactions. The effect of these is usually handled via the so-called Coulomb-correction~\cite{Nagy:2023zbg}. At the same time, the strong interaction of pions is usually negligible~\cite{Kincses:2019rug}, but is much more important for baryons and in case of non-identical correlations~\cite{Fabbietti:2020bfg,Kisiel:2014mma}.

%%%%%%%%%%%%%%%%%%%%%%%%%%%%%%%%%%%%%%%%%%
\section{Basics of Femtoscopy with Lévy Sources}\label{s:basicLevyHBT}

As mentioned above, a usual assumption for the shape of the (pair or single-particle) source is the Gaussian distribution. However, the observation of a long tail on the reconstructed source shape suggested~\cite{PHENIX:2007grx,PHENIX:2006nml} that for an adequate description, one can and should go beyond the Gaussian approximation. A plausible generalization is to utilize Lévy-stable distributions, as introduced in Ref.~\cite{Csorgo:2003uv} and utilized, for example, in Ref.~\cite{PHENIX:2017ino}. These appear when the conditions of the Central Limit Theorem are not met, particularly when the independent and identically distributed random variables (the sum of which we investigate) do not have a finite mean or variance. In this case, the Generalized Central Limit Theorem applies~\cite{Gnedenko:GCLT}, and the limiting distributions are the so-called Lévy-stable distributions~\cite{Nolan:Levy}. The spherically symmetric Lévy-stable distributions (or Lévy distributions for short) are defined as
\begin{align}\label{e:Levydef}
\mathcal{L}(\alpha,R,\mathbf{r})=\frac{1}{(2\pi)^3}\int d^3 \mathbf{q}\, e^{i\mathbf{q}\mathbf{r}} e^{-\frac{1}{2}|\mathbf{q}R|^{\alpha}},
\end{align}
where $\mathbf{r}$ is the spatial vector, the variable of the distribution, and $\alpha$ is the Lévy exponent, characterizing the shape of the distribution; the stability criterion (i.e., the limiting nature of the distribution) is met if $0<\alpha\leq 2$. Furthermore, $R$ is the spatial scale of the distribution, characterizing its size (even if the variance or root-mean-square is infinity) and corresponding to the HBT or femtoscopic radius in our case. Let us note here that if $\alpha<2$, then the distribution exhibits a power-law tail in $r=|\mathbf{r}|$, with
$\mathcal{L}(\alpha,R,\mathbf{r}) \propto (r/R)^{-(3+\alpha)}$.
Its angle-averaged version behaves as
$r^2 \mathcal{L}(\alpha,R,\mathbf{r}) \propto r^{-1-\alpha}$,
underlining the fact that these distributions have no finite variance for $\alpha<2$, and $r$ has no finite mean for $\alpha<1$ (clearly, the mean of $\mathbf{r}$ is zero, due to spherical symmetry).

Several (competing) physical phenomena may cause the appearance of Lévy distributions in high-energy physics:
\begin{itemize}
 \item Anomalous diffusion~\cite{Metzler:1999zz,Csanad:2007fr}, or Lévy-flight, probably due to hadronic rescattering; if this happens in the hadronic stage, then an ordering of the Lévy exponent is expected as
 $\alpha({\rm kaons})<\alpha({\rm pions})<\alpha({\rm protons})$, due to the cross-section and, hence, mean free path of the given particle species.~\cite{Csanad:2007fr}
 \item Jet fragmentation~\cite{Csorgo:2004sr}, where the fractal nature of the process creates the Lévy distribution; this may be a dominant effect in $e^+e^-$ collisions~\cite{L3:2011kzb}.
 \item Second-order phase transitions~\cite{Csorgo:2005it}, where the correlation length diverges near the critical point, and at the critical point, the spatial correlations exhibit a power-law tail with exponent $\eta$; this is one of the critical exponents, and its value is suggested to be $0.03631(3)$~\cite{El-Showk:2014dwa} for the 3D Ising model, or $0.50\pm0.05$ for the random external field 3D Ising model~\cite{Rieger:1995aa}---QCD is expected to be in the same universality class as one of them~\cite{Halasz:1998qr,Stephanov:1998dy}.
 \item Resonance decays~\cite{Csanad:2007fr,Kincses:2022eqq,Korodi:2022ohn}, where the power-law tail is generated by the set of resonances decaying into pions (or the given investigated particle species); this phenomenon is similar to the decay heat of used fuel rods in power plants, see Figure 1 of Ref.~\cite{Choi:2013aa}. Note that the simulations of Refs~\cite{Korodi:2022ohn,Kincses:2022eqq} indicate Lévy distributions in EPOS even before resonance decays; hence (at least in EPOS), these cannot be the only reason for the appearance of Lévy distributions.
\end{itemize}

\begin{figure}
    \includegraphics[width=0.49\linewidth]{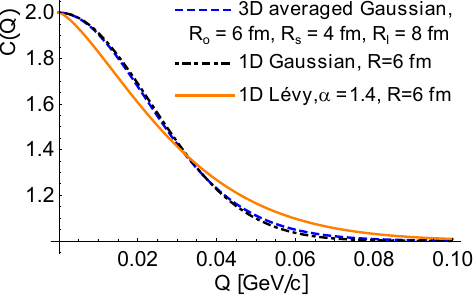} 
    \includegraphics[width=0.49\linewidth]{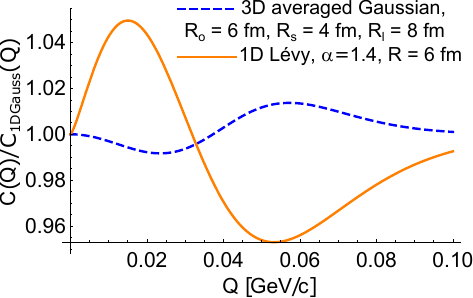}
    \caption{The left plot shows correlation functions $C(Q)$ (equivalent to $C_2(q)$ as denoted in the text) based on various source geometries: a 3D angle-averaged Gaussian with $R_{\rm out}=6$ fm, $R_{\rm side}=4$ fm, and $R_{\rm long}=8$ fm, a spherically symmetric (`1D') Gaussian with $R=6$ fm, and an also spherically symmetric (`1D') Lévy with $\alpha=1.4$ and $R=6$ fm. The right plot shows the ratio of the correlation functions based on a 3D Gaussian source and the spherically symmetric Lévy source, divided by the spherically symmetric Gaussian source.}
    \label{f:angle_averaging}
\end{figure}

Further ideas have been promoted recently, for example, event averaging and angular averaging, see Ref.~\cite{Tomasik:2019tjj,Cimerman:2019tku}. Note that in Refs.~\cite{Kincses:2022eqq,Korodi:2022ohn}, source shapes from the EPOS model have been analyzed on an event-by-event basis, and Lévy distributions appeared in individual events as well. Note furthermore that while a non-spherical Gaussian source is clearly exhibiting a non-Gaussian shape when measured in an angle-averaged fashion, a spherically symmetric Lévy distribution leads to quite different correlation functions, as illustrated in Figure~\ref{f:angle_averaging}. This shows that a 3D angle-averaged Gaussian (with \mbox{radii $R_{\rm out}=6$ fm}, $R_{\rm side}=4$ fm, and $R_{\rm long}=8$ fm in the Bertsch--Pratt frame~\cite{Pratt:1986cc,Bertsch:1989vn}) source leads to a non-Gaussian 1D correlation function. However, this apparent ``non-Gaussianity'' is distinctly different from a 1D (or spherically symmetric) correlation function based on a Lévy distribution. With respect to this, it shall also be noted that 3D HBT measurements in the longitudinally comoving system (LCMS) have been performed in PHENIX~\cite{Kurgyis:2018zck} and the resulting $\alpha$ exponent appeared to be the same as the one from the 1D analysis at~\cite{PHENIX:2017ino}, indicating that it is not non-sphericality that causes the appearance of Lévy distributions. Note that a further difference between 1D and 3D analyses is that, in 1D, one cannot easily make the transformation from the LCMS to the pair rest frame, where the Coulomb correction is calculated (the two-particle Schrödinger equation is a good approximation to the given problem only in this frame, where the nonrelativistic formulas may be utilized). However, in Ref.~\cite{Kurgyis:2020vbz}, simple but adequate approximation formulas were found, particularly for the source radius to be used in the Coulomb correction calculation.
 
%%%%%%%%%%%%%%%%%%%%%%%%%%%%%%%%%%%%%%%%%%
\section{Measures of the Source Extent}\label{s:hwhi_hwhm}

As discussed above, the source may exhibit various shapes, and even in the absence of a finite variance, one is interested in finding an ``assumption-independent'' measure of the source extent. One of these may be the half with at half maximum (HWHM), the distance (from the origin or center) where the probability density decreases to half of the maximal value (admitted at the origin or center usually). Another such measure may be the half width at half-integral, i.e., the radius of the region from where half of the particles are emitted. In the case of a Gaussian, the HWHM of the single-particle source can be readily calculated as
\vspace{-6pt}
\begin{align}
{\rm HWHM} = \sqrt{2\log(2)}\cdot R\approx 1.177\cdot R,
\end{align}
which is valid for both 1D and 3D Gaussian distributions with width $R$. The HWHI value for a 1D Gaussian is
 \vspace{-6pt}
\begin{align}
{\rm HWHI}_{\rm 1D\;Gauss} = \sqrt{2}\cdot {\rm erf}^{-1}\left(\frac{1}{2}\right)\cdot R \approx 0.664 \cdot R,
\end{align}
while for the 3D Gaussian, it can be given implicitly in terms of solving
\vspace{-6pt}
\begin{align}
{\rm erf}\left(\xi\right)+
\frac{2\xi}{\sqrt{\pi}} \exp\left(-\xi^2\right)=
\frac{1}{2},\quad\text{where}\quad\xi = \frac{{\rm HWHI}_{\rm 3D\;Gauss}}{\sqrt{2}R},
\end{align}
and the numerical value of the solution is
\vspace{-6pt}
\begin{align}
{\rm HWHI}_{\rm 3D\;Gauss}\approx1.538\cdot R.
\end{align}

\begin{figure}
    \centering
    \includegraphics[width=0.485\linewidth]{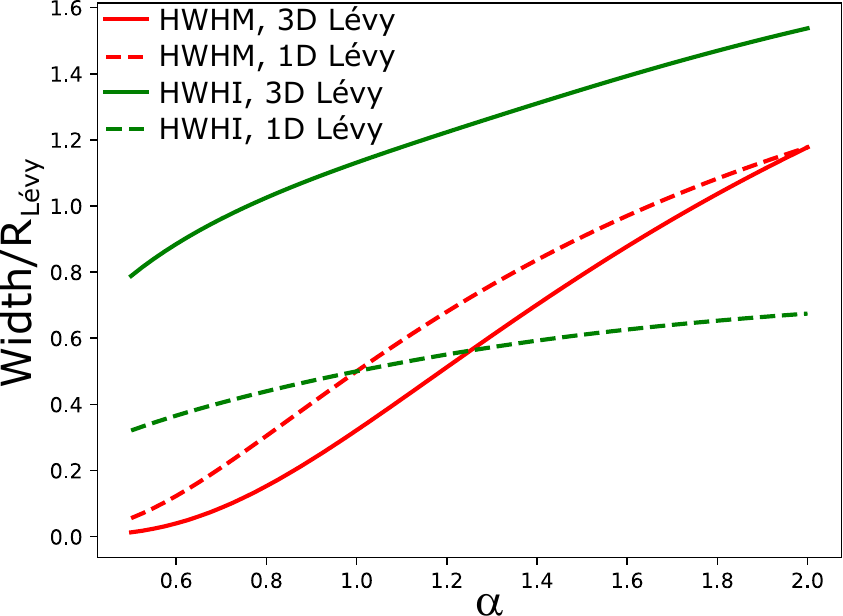}
    \includegraphics[width=0.505\linewidth]{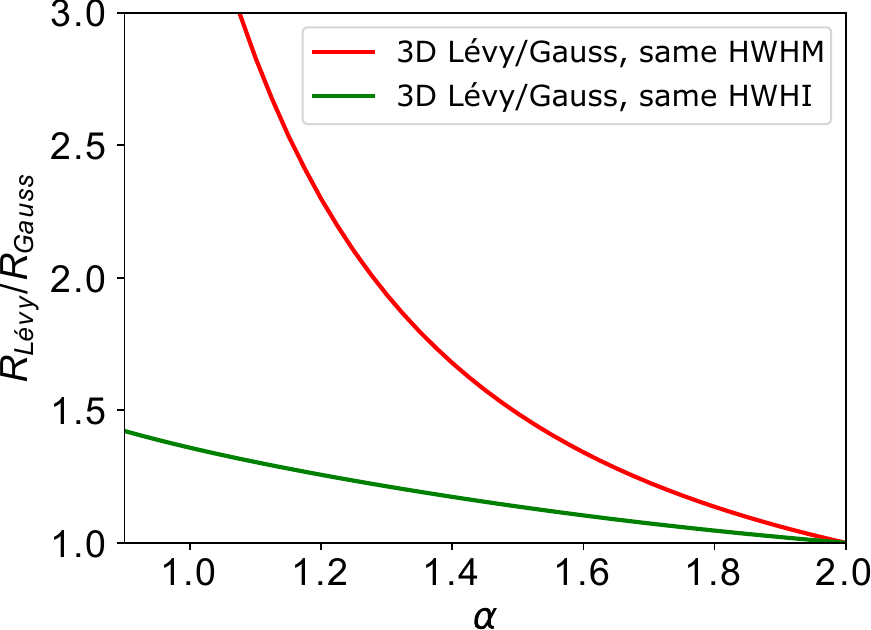}
    \caption{Left plot: two ``model-independent'' source measures, scaled by the L\'evy radius ($R_{\textnormal{L\'{e}vy}}$): half width at half maximum (HWHM, red lines) and half width at half integral (HWHI, green lines) are shown, for 1D and 3D sources (dashed and solid lines, respectively), as a function of the L\'evy-exponent $\alpha$. Right plot: ratio of 3D L\'evy and Gaussian radii when the HWMI (green) or the HWHM (red) is the same for the two distributions.}
    \label{f:hwhm_vs_whi}
\end{figure}

The above results can be understood for a Lévy distribution with $\alpha=2$. However, these values strongly depend on $\alpha$, as illustrated by the left plot of Figure~\ref{f:hwhm_vs_whi}. This means that either HWHI or HWHM can change due to either of the following:
\begin{itemize}
    \item A change in size (i.e., the $R_\textrm{Lévy}$ scale);
    \item A change in shape.
\end{itemize}

This means that if one is interested in a size change, then one may not be able to access this by investigating any of the ``assumption-independent'' measures (note furthermore that in order to experimentally access either HWHI or HWHM, one may need to extrapolate to large sizes, a region not accessible even via the so-called imaging technique~\cite{Brown:1997ku,PHENIX:2007grx}), but one needs to investigate the shape as well. This statement, of course, depends on what one regards as ``size'', but the above details underline the fact that any size measure may be entangled with the shape measure, and to extract size information, one may need to be able to quantify the shape of the source as well.

The right plot of Figure~\ref{f:hwhm_vs_whi} furthermore indicates the ratio of Lévy and Gaussian radii, when the HWHI or the HWMH of the two distributions is the same. This indicates that some sort of ``translation'' can be found between Lévy (free $\alpha$) and Gaussian ($\alpha=2$) radii: for example at $\alpha=1.3$, the Gaussian distribution with the same HWHM is of radius $R_{\rm 3D\;Gauss} \approx 1.94 R_{\alpha=1.3}$, while it has the same HWHI if $R_{\rm 3D\;Gauss} \approx 1.21 R_{\alpha=1.3}$.

%%%%%%%%%%%%%%%%%%%%%%%%%%%%%%%%%%%%%%%%%%
\section{Signs of the Critical Point in Femtoscopy}

As noted above, the Lévy exponent $\alpha$ was identified in Ref.~\cite{Csorgo:2005it} with the critical exponent of spatial correlations, usually denoted by $\eta$. It was furthermore suggested that, while far from the critical point, $\alpha$ may be of a value close to 2, closer to the critical point, or it may decrease to the above-mentioned values, close to or below 0.5. This behavior may be blurred by finite size effects; however, Refs.~\cite{Ballesteros:1996bd,Fytas:2023izo} found that the correlation exponent $\eta$ is largely unchanged by finite size effects in the 3D Ising model and also in the random field 3D Ising model. This suggests that it is not a priori impossible to search for the critical point via the measurement of $\alpha$. Furthermore, discussed, e.g., in Ref.~\cite{Lacey:2014wqa}, femtoscopic radii may be interpreted as signaling the vicinity of the critical point. Finally, one may note that the HBT radii are also connected to the lifetime of the quark medium~\cite{Makhlin:1987gm}, which is also influenced by the equation of state of the medium. Hence, the collision energy dependence of HBT radii serves as an important tool to investigate the QCD phase diagram.

\begin{figure}
    \centering
    \includegraphics[width=0.75\linewidth]{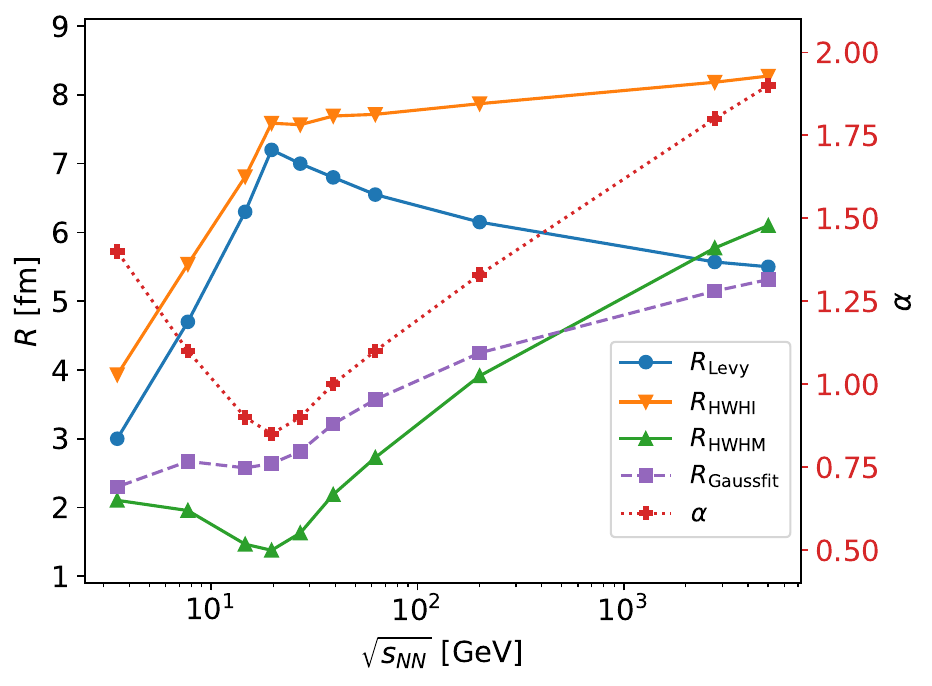}
    \caption{Various source geometry characteristics versus collision energy ($\sqrt{s_{_{NN}}}$) in a hypothetical scenario where $\alpha$ (red crosses, right vertical axis) has a minimum and $R_{\rm Levy}$ a maximum near $\sqrt{s_{_{NN}}}=20$ GeV. The following calculated source measures were calculated and are shown: half width at half integral ($R_{\rm HWHI}$, orange downward-facing triangles), half width at half maximum ($R_{\rm HWHM}$, green upward-facing triangles), and a radius from a forced Gaussian fit ($R_{\rm Gaussfit}$, purple squares).}
    \label{f:levy_Gauss_R_vs_energy}
\end{figure}

Let us now suppose a scenario where the Lévy scale $R_\textrm{Lévy}$ exhibits a maximum near a collision energy of \mbox{$\sqrt{s_{_{NN}}}=20$ GeV}, while the Lévy exponent $\alpha$ has a minimum at the same location, as shown in Figure~\ref{f:levy_Gauss_R_vs_energy}. The values depicted on this plot are purely hypothetical, assumed to take the value shown on the figure for illustration purposes. Let us then generate correlation functions with a toy model where the supposed Lévy parameters are generated with realistic statistical uncertainties (following those from experimental data, such as the one in Ref.~\cite{PHENIX:2017ino}). We investigate three different size measures (besides $R_\textrm{Lévy}$, which is an input, but it was confirmed that from a Lévy fit it can be retained):
\begin{itemize}
 \item $R_{\rm Gauss\;fit}$: This is obtained via fitting a correlation function to the data (generated via the simple toy model), with the assumption of $\alpha=2$.
 \item HWHI: half width at half-integral, as discussed in Section~\ref{s:hwhi_hwhm}.
 \item HWHM: half width at half maximum, as discussed in Section~\ref{s:hwhi_hwhm}.
\end{itemize}

Figure~\ref{f:levy_Gauss_R_vs_energy} shows the results, and the observation is that $R_{\rm Gauss\;fit}$, contrary to the input $R_\textrm{Lévy}$, shows an approximately monotonic increase with $\sqrt{s_{_{NN}}}$. On the other hand, there is a minimum in HWHM, and HWHI exhibits a trend change.

These findings again underline the importance of quantifying the shape and the scale (size) of the source at the same time, as these two are entangled in source size measures such as HWHI, HWHM, or Gaussian radius.

%%%%%%%%%%%%%%%%%%%%%%%%%%%%%%%%%%%%%%%%%%
\section{Experimental Results from SPS through RHIC to LHC}
%Published and preliminary results (already published in proceedings)

After the above phenomenological considerations, let us discuss recent experimental results as well. These can be compared to the first thorough two-pion Lévy HBT analysis in heavy-ion collisions, as detailed in Ref.~\cite{PHENIX:2017ino}. In this paper, $\sqrt{s_{_{NN}}}=200$ GeV Au+Au collisions with 0--30\% centrality were analyzed, and $\alpha$ values around 1.2 were found, with a close to negligible dependence on $m_T$. These were later confirmed in a preliminary 3D analysis~\cite{Kurgyis:2018zck}. In both analyses~\cite{PHENIX:2017ino,Kurgyis:2018zck}, a decreasing trend of the Lévy HBT scale $R$ was found with $m_T$, a feature usually attributed to the expansion of the fireball, as this leads to smaller sources for particles with larger momenta~\cite{Makhlin:1987gm,Csorgo:1995bi,Lisa:2008gf}. This, however, was calculated and predicted for Gaussian HBT radii. Hence, it is an open question why this feature of $R\propto 1/\sqrt{m_T}$ holds for Lévy HBT radii as well. In the mentioned analysis of Ref.~\cite{PHENIX:2017ino}, an interesting behavior of the correlation strength parameter $\lambda$ was found: it exhibits a ``hole'' (i.e., a strong decrease) in the small-$m_T$ region, attributed to an in-medium mass decrease of the $\eta'$ meson (see Ref.~\cite{PHENIX:2017ino} for details). These are the features of the Lévy HBT parameters ($\alpha$, $R$, $\lambda$) that shall be investigated in recent results as well.

\begin{figure}
    \centering
    \includegraphics[width=0.75\linewidth]{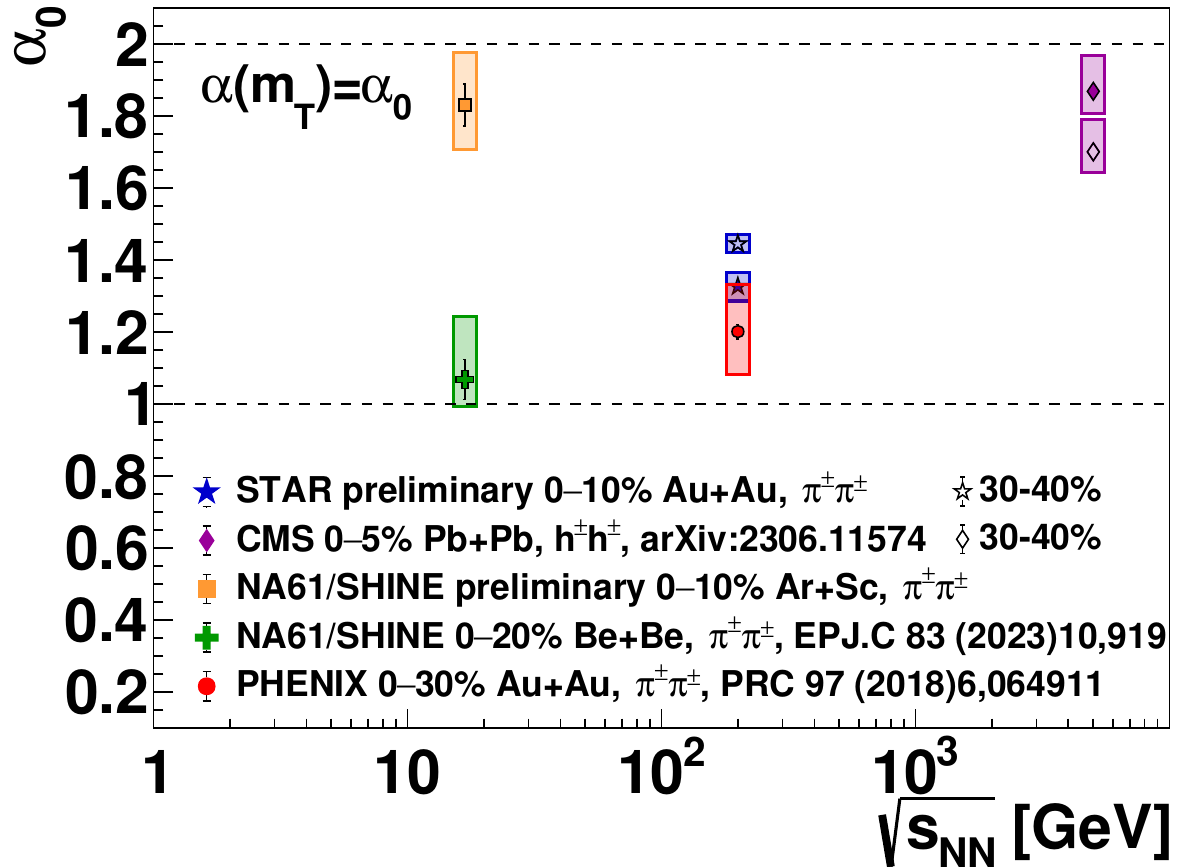}
    \caption{Experimental results on the mean ($m_T$-averaged) $\alpha$ value (denoted here as $\alpha_0$), as a function of collision energy ($\sqrt{s_{_{NN}}}$). The STAR preliminary results are from~\cite{Kincses:ISMD23}. The CMS $\sqrt{s_{_{NN}}} = 5.02$ TeV Pb+Pb results are from Ref.~\cite{CMS:2023xyd}. The PHENIX $\sqrt{s_{_{NN}}} = 200$ GeV 0--30\% Au+Au result is from Ref.~\cite{PHENIX:2017ino}. The 150$A$ GeV ($\sqrt{s_{_{NN}}} = 16.82$) Ar+Sc results are from Ref.~\cite{Porfy:2023yii}, while the 150$A$ GeV Be+Be results are from Ref.~\cite{NA61SHINE:2023qzr}.}
    \label{f:alpha0_sqrts_allexp}
\end{figure}

The detailed centrality and average transverse mass dependence of these two-pion Lévy HBT parameters have been measured in Au+Au collisions with STAR at \mbox{$\sqrt{s_{_{NN}}}=200$ GeV} and reported recently by D. Kincses at the 52nd International Symposium on Multiparticle Dynamics and the XVI. Workshop of Particle Correlations and Femtoscopy, and in Ref.~\cite{Kincses:ISMD23}.
%%REF TO BE ADDED IF AVAILABLE ON ARXIV!!!
These results are summarized in the same Special Issue as this paper. We note here that a similar $\alpha$ value was found as in Ref.~\cite{PHENIX:2017ino}, as shown in the summary plot of Figure~\ref{f:alpha0_sqrts_allexp}, and similar features for the $m_T$-dependence of $R$ and $\lambda$ parameters were found as in Ref.~\cite{PHENIX:2017ino}.

A similar analysis in the CMS experiment in $\sqrt{s_{_{NN}}}=5.02$ TeV Pb+Pb collisions was also performed (albeit without particle identification, under the assumption that most detected particles are pions), as reported in Ref.~\cite{CMS:2023xyd}. Here, it was found that the Lévy scale $R$ follows a similar trend versus $m_T$ as present in the RHIC results discussed above. It was also found that the exponent $\alpha$ is closer to the Gaussian case but clearly not equal to it. Furthermore, $\alpha$ values were found to be larger (closer to 2) for central collisions; this behavior is distinctly different from what was found in STAR. Figure~\ref{f:alpha0_sqrts_allexp} also shows a central and a peripheral $\alpha$ measurement for both STAR and CMS, where the opposite trend \mbox{is apparent.} %!!! %We concentrate here, however, on the collision energy dependence, hence showing the 0-5\% centrality results in Figure~\ref{f:alpha0_sqrts_allexp}.

A two-pion HBT measurement was reported by the NA61/SHINE experiment in Refs.~\cite{NA61SHINE:2023qzr,Porfy:2023yii} in 150$A$ GeV ($\sqrt{s_{_{NN}}}=16.82$ GeV) Be+Be and Ar+Sc collisions. For the Lévy index $\alpha$, a significant difference between the two systems was found, a result yet to be understood. Average $\alpha$ values for both systems are shown in Figure~\ref{f:alpha0_sqrts_allexp}. For the Lévy scale $R$, the difference between the two systems is close to the initial size ratio of roughly 1.6 (based on nuclear sizes), and it decreases with $m_T$, similarly to the case of RHIC and LHC. Interestingly, however, the $\lambda$ parameter does not show the ``hole'' mentioned above for the RHIC case, which may be a signal of the in-medium mass modification effect being turned off at SPS energies.

As mentioned, Figure~\ref{f:alpha0_sqrts_allexp} compares the mean $\alpha$ values at several collision energies from SPS through RHIC to LHC. Interestingly, the 5.02 TeV Pb+Pb results and the \mbox{16.82 GeV} Ar+Sc results yield the largest $\alpha$ values, while those for 200 GeV Au+Au and 16.82 GeV Be+Be collisions are significantly smaller. It is important to note, however, that the kinematic range of these experiments is also quite different: While there is essentially no low-momentum cut in NA61/SHINE (as in the fixed-target setup, small momentum particles can also be detected), there is one of about 0.2 GeV (in transverse momentum) in the RHIC experiments, while in case of CMS, particles with transverse momenta smaller than 0.5 GeV are discarded. This may also strongly influence the results of Figure~\ref{f:alpha0_sqrts_allexp}.

Besides the pion parameters, kaon analyses have been performed both with PHENIX~\cite{Kovacs:2023qax} and with STAR~\cite{Mukherjee:2023hrz} in central $\sqrt{s_{_{NN}}}=200$ GeV Au+Au collisions. Contrary to the anomalous diffusion expectation of $\alpha({\rm kaons})<\alpha({\rm pions})$, as detailed in Section~\ref{s:basicLevyHBT}, it was found that the $\alpha$ parameter for kaons is close to or slightly above to the one for pions. It was also found that the Lévy radii ($R$) for kaons and pions are compatible at the same $m_T$. These results indicate that it is not just hadronic anomalous diffusion that is responsible for the apparent Lévy source distributions, and that further modeling is required to understand the processes shaping \mbox{the source.}

\begin{figure}
    \centering
    \includegraphics[width=0.75\linewidth]{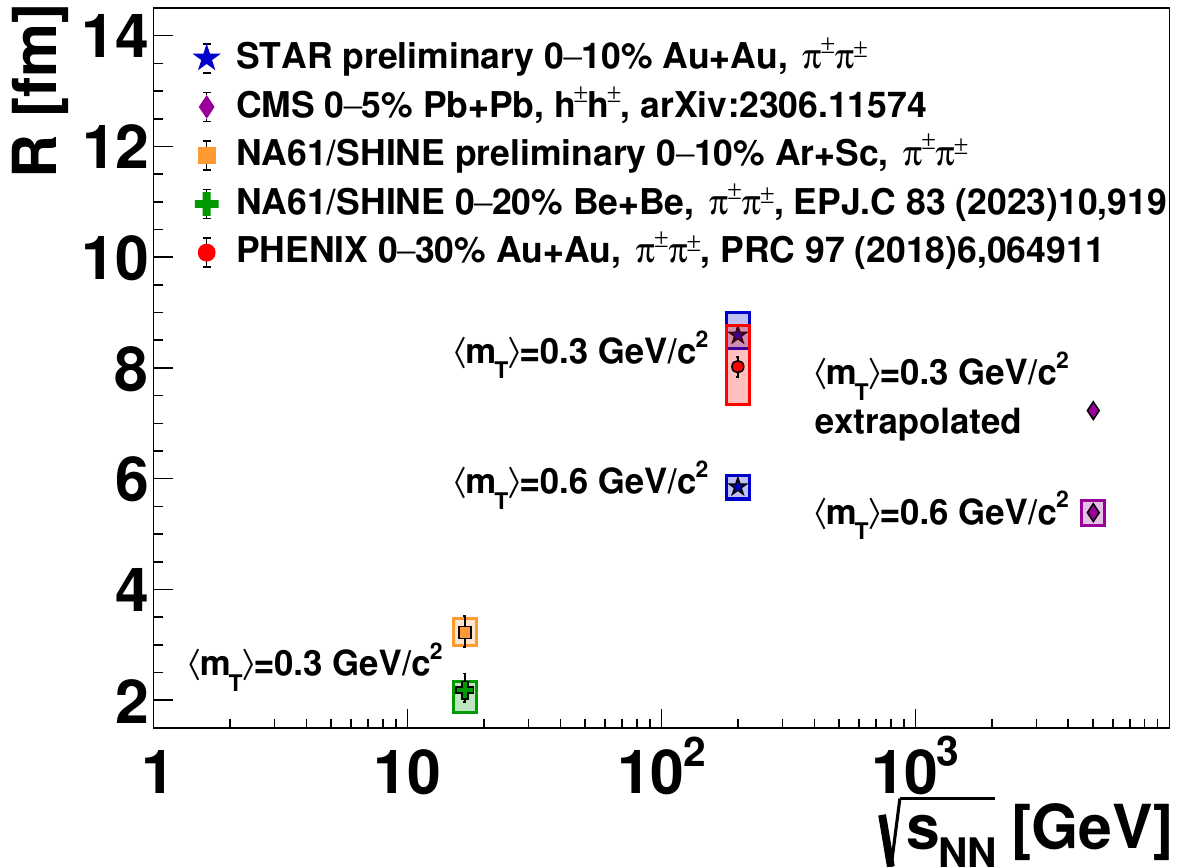}
    \caption{Experimental results on the $R$ value (at a mean $m_T$ of 0.3 and 0.6 GeV/$c^2$), as a function of collision energy ($\sqrt{s_{_{NN}}}$). The source of the data is the same as for Figure~\ref{f:alpha0_sqrts_allexp}. The CMS data point for $m_T=0.3$ GeV/$c^2$ was obtained via an extrapolation, as noted in the text.}
    \label{f:R_sqrts_allexp}
\end{figure}

The $R$ parameter obtained in the above-mentioned experimental analyses exhibits a typical and characteristic dependence on $m_T$: it decreases significantly as $m_T$ increases, due to the non-static nature of the source and the fact that $R$ is not a geometric size but a kind of measure of homogeneity, as mentioned in the introduction. Hence, $m_T$-averaging would not be meaningful. To facilitate a comparison, in Figure~\ref{f:R_sqrts_allexp}, we show $R$ values from SPS~\cite{NA61SHINE:2023qzr,Porfy:2023yii} and RHIC~\cite{PHENIX:2017ino,Kincses:ISMD23} at $m_T=0.3$ GeV/$c^2$, along with an extrapolation of the CMS data~\cite{CMS:2023xyd} to this $m_T$ value, using the formula for $R(m_T)$ utilized in Ref.~\cite{Kincses:ISMD23}. We furthermore compare RHIC and LHC results at $m_T=0.6$ GeV/$c^2$ (extrapolation of the SPS data to 0.6~GeV/$c^2$ is less feasible due to the larger uncertainties there). It is clear that $R$ decreases with $m_T$, as detailed in the above-cited experimental papers as well. It is also clear that the Lévy scale depends on the initial system size; in particular, $R$ for Be+Be is smaller than for Ar+Sc at the same energy. There is also an increase from SPS to RHIC, partly probably due to the system size change but also due to the increased energy. Rather surprisingly, however, Lévy scales at the LHC are not larger than those at RHIC but slightly smaller. This may be connected to the larger $\alpha$ measured at LHC and the anti-correlation of $R$ and $\alpha$, as demonstrated by Figure~7c of Ref.~\cite{PHENIX:2017ino}, Figure~5 of Ref.~\cite{Kincses:2022eqq}, and Figures~4 and 5 of Ref.~\cite{Korodi:2022ohn}.

The third important physical parameter of HBT correlation functions is the correlation strength parameter $\lambda$, as introduced in the first section. It was observed at RHIC~\cite{PHENIX:2017ino} that there is a decrease of $\lambda(m_T)$ at small $m_T$ values. The core--halo model interpretation utilized in Ref.~\cite{PHENIX:2017ino} suggests that this may indirectly connect to an in-medium mass modification of the $\eta'$ meson. It is, hence, quite interesting that this behavior is absent at SPS, as if this effect could be ``turned off'' when decreasing the collision energy from RHIC to SPS. Investigating this (the $m_T$-dependence of $\lambda$) at various energies also in data from the RHIC beam energy scan will be important and interesting. Note that no particle identification was employed in the CMS analysis of Ref.~\cite{CMS:2023xyd}. Hence, $\lambda$ cannot be interpreted in terms of the core--halo model only. In Ref.~\cite{CMS:2023xyd}, an attempt was made to correct for the lack of particle identification, and a new parameter $\lambda^*$ was shown. However, the accessible $m_T$-range of CMS does not allow for investigating the phenomenon mentioned above (i.e., the decrease of $\lambda$ at small $m_T$).

%%%%%%%%%%%%%%%%%%%%%%%%%%%%%%%%%%%%%%%%%%
\section{Discussion and Conclusions}

We have detailed several phenomenological and experimental results related to femtoscopy with Lévy sources. In light of these, the natural question arises: is it advantageous to utilize Lévy distributed sources in describing momentum correlations? We have shown above that the appearing extra parameter, the Lévy exponent $\alpha$ has a physical meaning for each of the possible physical reasons for its appearance (jet fragmentation, critical phenomena, anomalous diffusion, resonance decays). In order to interpret it, one needs to disentangle the above effects. Equally importantly, when measuring source size ($R$) or strength ($\lambda$) parameters with a Gaussian assumption, the actual size and shape information are entangled, and the more general Lévy assumption leads to refined results. One shall also note that the source, of course, does not necessarily exhibit an exact Lévy shape---deviations from it can be investigated via the method of Lévy expansion~\cite{Novak:2016cyc}, and in Ref.~\cite{PHENIX:2017ino} all expansion terms turned out to be zero, i.e., no deviation from the Lévy shape was found. This is not necessarily true, however, for all other experiments, and a similar expansion analysis can be performed in those cases as well.

One may also note that there are a few complications when fitting correlation functions based on Lévy sources. First of all, the extra parameter ($\alpha$) is significantly correlated with the other physical parameters ($R$ and $\lambda$). Hence, fit range stability and statistical uncertainties need a careful approach. Furthermore, the Coulomb correction is more complicated to determine in this case; but as noted above, a novel method exits~\cite{Nagy:2023zbg} with which this can be treated in a self-consistent and accurate calculation.

It is furthermore an important question whether utilizing the imaging technique mentioned above and detailed in Ref.~\cite{Brown:1997ku} or the related technique of spherical decomposition~\cite{Kisiel:2009iw} makes the Lévy source assumption superfluous. Indeed, these methods help us map out the particle emitting source. In order to quantify the collision energy or centrality evolution of the source (for example, its size), one then would proceed to determine source measures such as HWHI and HWHM (or even a root-mean-square)---or fit the imaged sources with a suitable shape, such as a Gaussian or a Lévy distribution. If this shape indeed describes the data, then it is equivalent to fitting the source or the correlation function. However, in the case of the correlation function, the uncertainties are more clear; hence, even in that case, it is advantageous to fit this experimental object. Therefore, the utilization of the Lévy sources (possibly with the Lévy expansion technique) has no real disadvantage compared to imaging as well.

In summary, Lévy sources have been observed from SPS through RHIC to LHC. The possible reasons for the appearance of Lévy sources in high-energy physics include, depending on system size and collision energy, jet fragmentation, critical phenomena, resonance decays, and anomalous diffusion, which need to be disentangled. In measurements, Lévy exponent values between 1 and 2 were found, with a nontrivial collision energy and system size dependence, suggesting that multiple phenomena contribute to the observed $\alpha$. The particle type dependence was also found to be nontrivial, not following the simplest predictions based on anomalous diffusion in the hadronic stage. The Lévy scale (or HBT radius) $R$ was found to exhibit an $m_T$-dependence that was predicted for Gaussian radii: a feature yet to be understood via model calculations. Finally, femtoscopy with Lévy sources provides a method to measure the correlation strength $\lambda$ accurately, and differences were found in the low-$m_T$ behavior between RHIC and SPS, possibly hinting at an effect being absent at SPS energies. These results underline the importance of considering Lévy shapes when describing momentum correlation functions in high-energy physics and the need for more detailed phenomenological studies to explore all factors contributing to the shape of the particle emitting source.

\section*{Acknowledgments}
We acknowledge the support of NKFIH grants K-138136, TKP2021-NKTA-64, K-146913, and PD-146589. We also thank M. Nagy for insightful discussions.


\begin{thebibliography}{55}

\bibitem{Lednicky:2001qv}
R.~Lednicky, \emph{{Femtoscopy with unlike particles}}, in \emph{{International
  Workshop on the Physics of the Quark Gluon Plasma}} (2001), \texttt
  arXiv:{nucl-th/0112011}

\bibitem{Lisa:2005dd}
M.A. Lisa, S.~Pratt, R.~Soltz, U.~Wiedemann, Ann. Rev. Nucl. Part. Sci.
  \textbf{55}, 357 (2005), \texttt arXiv:{nucl-ex/0505014}

\bibitem{HanburyBrown:1952na}
R.~Hanbury~Brown, R.C. Jennison, D.G.M. K., Nature \textbf{170}, 1061–1063
  (1952)

\bibitem{Goldhaber:1960sf}
G.~Goldhaber, S.~Goldhaber, W.Y. Lee, A.~Pais, Phys. Rev. \textbf{120}, 300
  (1960)

\bibitem{Csanad:2018vgk}
M.~Csanád [for the PHENIX Collaboration], J. Phys. Conf. Ser. \textbf{1070}, 012026 (2018), \texttt
  arXiv:{1806.05745}

\bibitem{PHENIX:2017ino}
A.~Adare et~al. [PHENIX Collaboration], Phys. Rev. C \textbf{97}, 064911 (2018), [Erratum:
  Phys.Rev.C 108, 049905 (2023)], \texttt arXiv:{1709.05649}

\bibitem{Kurgyis:2020vbz}
B.~Kurgyis, D.~Kincses, M.~Nagy, M.~Csanád, Universe \textbf{9}, 328 (2023),
  \texttt arXiv:{2007.10173}

\bibitem{Nagy:2023zbg}
M.~Nagy, A.~Purzsa, M.~Csanád, D.~Kincses, Eur. Phys. J. C \textbf{83}, 1015
  (2023), \texttt arXiv:{2308.10745}

\bibitem{Korodi:2022ohn}
B.~K\'orodi, D.~Kincses, M.~Csanád, Phys. Lett. B \textbf{847}, 138295
  (2023), \texttt arXiv:{2212.02980}

\bibitem{Csorgo:1994in}
T.~Csörgő, B.~Lörstad, J.~Zimányi, Z. Phys. C \textbf{71}, 491 (1996), \texttt
  arXiv:{hep-ph/9411307}

\bibitem{Csorgo:1999sj}
T.~Csörgő, Acta Phys. Hung. A \textbf{15}, 1 (2002), \texttt
  arXiv:{hep-ph/0001233}

\bibitem{Csanad:2020qtu}
M.~Csanád, A.~Jakovác, S.~Lökös, A.~Mukherjee, S.K. Tripathy,
  \emph{{Multi-particle quantum-statistical correlation functions in a
  Hubble-expanding hadron gas}} (World Scientific, 2020), pp. 261--273, \texttt
  arXiv:{2007.07167},
  \texttt{{https://www.worldscientific.com/doi/abs/10.1142/9789811238406\_0023}}

\bibitem{STAR:2020dav}
J.~Adam et~al. [STAR Collaboration], Phys. Rev. C \textbf{103}, 034908 (2021), \texttt
  arXiv:{2007.14005}

\bibitem{Lacey:2014wqa}
R.A. Lacey, Phys. Rev. Lett. \textbf{114}, 142301 (2015), \texttt
  arXiv:{1411.7931}

\bibitem{STAR:2014shf}
L.~Adamczyk et~al. [STAR Collaboration], Phys. Rev. C \textbf{92}, 014904 (2015), \texttt
  arXiv:{1403.4972}

\bibitem{PHENIX:2014pnh}
A.~Adare et~al. [PHENIX Collaboration], Phys. Rev. Lett. \textbf{112}, 222301 (2014), \texttt
  arXiv:{1401.7680}

\bibitem{Lokos:2016fze}
S.~L\"ok\"os, M.~Csanád, B.~Tom\'a\v{s}ik, T.~Cs\"org\H{o}, Eur. Phys. J. A
  \textbf{52}, 311 (2016), \texttt arXiv:{1604.07470}

\bibitem{Cimerman:2017lmm}
J.~Cimerman, B.~Tomasik, M.~Csanád, S.~Lökös, Eur. Phys. J. A \textbf{53}, 161
  (2017), \texttt arXiv:{1702.01735}

\bibitem{Kincses:2019rug}
D.~Kincses, M.I. Nagy, M.~Csanád, Phys. Rev. C \textbf{102}, 064912 (2020),
  \texttt arXiv:{1912.01381}

\bibitem{Fabbietti:2020bfg}
L.~Fabbietti, V.~Mantovani~Sarti, O.~Vazquez~Doce, Ann. Rev. Nucl. Part. Sci.
  \textbf{71}, 377 (2021), \texttt arXiv:{2012.09806}

\bibitem{Kisiel:2014mma}
A.~Kisiel, H.~Zbroszczyk, M.~Szyma\'nski, Phys. Rev. C \textbf{89}, 054916
  (2014), \texttt arXiv:{1403.0433}

\bibitem{PHENIX:2007grx}
S.~Afanasiev et~al. [PHENIX Collaboration], Phys. Rev. Lett. \textbf{100}, 232301 (2008),
  \texttt arXiv:{0712.4372}

\bibitem{PHENIX:2006nml}
S.S. Adler et~al. [PHENIX Collaboration], Phys. Rev. Lett. \textbf{98}, 132301 (2007),
  \texttt arXiv:{nucl-ex/0605032}

\bibitem{Csorgo:2003uv}
T.~Csörgő, S.~Hegyi, W.A. Zajc, Eur. Phys. J. C \textbf{36}, 67 (2004), \texttt
  arXiv:{nucl-th/0310042}

\bibitem{Gnedenko:GCLT}
B.V. Gnedenko, A.N. Kolmogorov, K.L. Chung, \emph{Limit Distributions For Sums
  Of Independent Random Variables} (Addisson-Wesley, Cambridge, MA, 1954)

\bibitem{Nolan:Levy}
J.P. Nolan, \emph{Univariate Stable Distributions}, Springer Series in
  Operations Research and Financial Engineering (Springer, New York, NY, 2020),
  \texttt{https://doi.org/10.1007/978-3-030-52915-4}

\bibitem{Metzler:1999zz}
R.~Metzler, E.~Barkai, J.~Klafter, Phys. Rev. Lett. \textbf{82}, 3563 (1999)

\bibitem{Csanad:2007fr}
M.~Csanád, T.~Csörgő, M.~Nagy, Braz. J. Phys. \textbf{37}, 1002 (2007), \texttt
  arXiv:{hep-ph/0702032}

\bibitem{Csorgo:2004sr}
T.~Csörgő, S.~Hegyi, T.~Novák, W.A. Zajc, Acta Phys. Polon. B \textbf{36}, 329
  (2005), \texttt arXiv:{hep-ph/0412243}

\bibitem{L3:2011kzb}
P.~Achard et~al. [L3 Collaboration], Eur. Phys. J. C \textbf{71}, 1648 (2011), \texttt
  arXiv:{1105.4788}

\bibitem{Csorgo:2005it}
T.~Csörgő, S.~Hegyi, T.~Novák, W.A. Zajc, AIP Conf. Proc. \textbf{828}, 525
  (2006), \texttt arXiv:{nucl-th/0512060}

\bibitem{El-Showk:2014dwa}
S.~El-Showk, M.F. Paulos, D.~Poland, S.~Rychkov, D.~Simmons-Duffin, A.~Vichi,
  J. Stat. Phys. \textbf{157}, 869 (2014), \texttt arXiv:{1403.4545}

\bibitem{Rieger:1995aa}
H.~Rieger, Phys. Rev. B \textbf{52}, 6659 (1995)

\bibitem{Halasz:1998qr}
A.M. Halasz, A.D. Jackson, R.E. Shrock, M.A. Stephanov, J.J.M. Verbaarschot,
  Phys. Rev. D \textbf{58}, 096007 (1998), \texttt arXiv:{hep-ph/9804290}

\bibitem{Stephanov:1998dy}
M.A. Stephanov, K.~Rajagopal, E.V. Shuryak, Phys. Rev. Lett. \textbf{81}, 4816
  (1998), \texttt arXiv:{hep-ph/9806219}

\bibitem{Kincses:2022eqq}
D.~Kincses, M.~Stefaniak, M.~Csanád, Entropy \textbf{24}, 308 (2022), \texttt
  arXiv:{2201.07962}

\bibitem{Choi:2013aa}
H.J. Choi, J.Y. Lee, J.~Choi, Nuclear Engineering and Technology \textbf{45},
  29 (2013)

\bibitem{Tomasik:2019tjj}
B.~Tom\'a\v{s}ik, J.~Cimerman, C.~Plumberg, Universe \textbf{5}, 148 (2019)

\bibitem{Cimerman:2019tku}
J.~Cimerman, B.~Tom\'a\v{s}ik, C.~Plumberg, Phys. Part. Nucl. \textbf{51}, 3
  (2020), \texttt arXiv:{1909.00278}

\bibitem{Pratt:1986cc}
S.~Pratt, Phys. Rev. D \textbf{33}, 1314 (1986)

\bibitem{Bertsch:1989vn}
G.F. Bertsch, Nucl. Phys. A \textbf{498}, 173C (1989)

\bibitem{Kurgyis:2018zck}
B.~Kurgyis [for the PHENIX Collaboration], \emph{{Three dimensional L\'evy HBT results from PHENIX}},
  in \emph{{13th Workshop on Particle Correlations and Femtoscopy}} (2018),
  \texttt arXiv:{1809.09392}

\bibitem{Brown:1997ku}
D.A. Brown, P.~Danielewicz, Phys. Lett. B \textbf{398}, 252 (1997), \texttt
  arXiv:{nucl-th/9701010}

\bibitem{Ballesteros:1996bd}
H.G. Ballesteros, L.A. Fernandez, V.~Martin-Mayor, A.~Munoz~Sudupe, Phys. Lett.
  B \textbf{387}, 125 (1996), \texttt arXiv:{cond-mat/9606203}

\bibitem{Fytas:2023izo}
N.G. Fytas, V.~Mart\'\i{}n-Mayor, G.~Parisi, M.~Picco, N.~Sourlas, Phys. Rev. E
  \textbf{108}, 044146 (2023), \texttt arXiv:{2307.01809}

\bibitem{Makhlin:1987gm}
A.N. Makhlin, Y.M. Sinyukov, Z. Phys. C \textbf{39}, 69 (1988)

\bibitem{Csorgo:1995bi}
T.~Csörgő, B.~Lörstad, Phys. Rev. C \textbf{54}, 1390 (1996), \texttt
  arXiv:{hep-ph/9509213}

\bibitem{Lisa:2008gf}
M.A. Lisa, S.~Pratt, \emph{{Femtoscopically Probing the Freeze-out
  Configuration in Heavy Ion Collisions}} (Springer-Verlag Berlin Heidelberg,
  2010), chap.~21, \texttt arXiv:{arXiv:0811.1352}

\bibitem{Kincses:ISMD23}
Kincses, D. [for the STAR Collaboration], \emph{Pion interferometry with Levy sources in $\sqrt{s_\mathrm{NN}}$ = 200
  GeV Au+Au Collisions at STAR} in \emph{Proceedings of the ISMD 2023 Conference, Gyöngyös, Hungary, 20--26 August 2023}.

\bibitem{CMS:2023xyd}
A.~Tumasyan et~al. [CMS Collaboration], \texttt arXiv:{2306.11574}

\bibitem{NA61SHINE:2023qzr}
H.~Adhikary et~al. [NA61/SHINE Collaboration], Eur. Phys. J. C \textbf{83}, 919 (2023),
  \texttt arXiv:{2302.04593}

\bibitem{Porfy:2023yii}
B.~Pórfy [for the NA61/SHINE Collaboration], Universe \textbf{9}, 298 (2023), \texttt
  arXiv:{2306.08696}

\bibitem{Kovacs:2023qax}
L.~Kov\'acs, Universe \textbf{9}, 336 (2023), \texttt arXiv:{2307.09573}

\bibitem{Mukherjee:2023hrz}
A.~Mukherjee, Universe \textbf{9}, 300 (2023), \texttt arXiv:{2306.13668}

\bibitem{Novak:2016cyc}
T.~Nov\'ak, T.~Cs\"org\H{o}, H.C. Eggers, M.~de~Kock, Acta Phys. Polon. Supp.
  \textbf{9}, 289 (2016), \texttt arXiv:{1604.05513}

\bibitem{Kisiel:2009iw}
A.~Kisiel, D.A. Brown, Phys. Rev. C \textbf{80}, 064911 (2009), \texttt
  arXiv:{0901.3527}

\end{thebibliography}
\end{document}